\documentclass{article}

\usepackage{pdflscape}
\usepackage{makeidx}
\usepackage{authblk}
\usepackage{hyperref}
\usepackage{graphicx}
\usepackage{caption}
\usepackage{subcaption}
\usepackage{color}
\usepackage[usenames,dvipsnames,svgnames,table]{xcolor}
\usepackage{booktabs}
\usepackage{threeparttable}
\usepackage{appendix}
\usepackage{textcomp}
\usepackage{gensymb}
\usepackage[round]{natbib}
\usepackage{csquotes}
\usepackage{mathtools}
\usepackage{xfrac}
\usepackage{enumitem}
\hypersetup{
backref=true, 
bookmarks=true, 
unicode=false, 
pdftoolbar=true, 
pdfmenubar=true, 
pdffitwindow=false, 
pdfstartview={FitH}, 
pdftitle={}, 
pdfauthor={}, 
pdfsubject={}, 
pdfkeywords={}{}, 
pdfnewwindow=true, 
colorlinks=true, 
linkcolor=BlueViolet, 
citecolor=NavyBlue, 
urlcolor=blue,
}

\title{Population and trends in the global mean temperature}
\author[a b c d e]{Richard S.J. Tol}
\affil[a]{Department of Economics, University of Sussex, BN1 9SL Falmer, United Kingdom, r.tol@sussex.ac.uk}
\affil[b]{Institute for Environmental Studies, Vrije Universiteit Amsterdam, The Netherlands}
\affil[c]{Department of Spatial Economics, Vrije Universiteit Amsterdam, The Netherlands}
\affil[d]{Tinbergen Institute, Amsterdam, The Netherlands}
\affil[e]{CESifo, Munich, Germany}

\date{\today}

\begin{document}

\maketitle

\begin{abstract}
The Fisher Ideal index, developed to measure price inflation, is applied to define a population-weighted temperature trend. This method has the advantages that the trend is representative for the population distribution throughout the sample but without conflating the trend in the population distribution and the trend in the temperature. I show that the trend in the global area-weighted average surface air temperature is different in key details from the population-weighted trend. I extend the index to include urbanization and the urban heat island effect. This substantially changes the trend again. I further extend the index to include international migration, but this has a minor impact on the trend.
\end{abstract}



\section{Introduction}
The global annual mean surface air temperature is typically defined as an area-weighted average of local temperatures. This definition makes sense from a meteorological perspective, but the distribution of the human population over the surface of the planet is rather uneven. Hardly any humans live at sea, and desert, tundra and rainforest show low population densities. Climate change in such areas is much less relevant to the human condition than warming in cities. In this paper, I therefore discuss temperature trends as experienced by humans. I do so in three steps, accounting for changes in population density, for urbanization, and for migration.

Unlike area, population changes over time. Population-weights therefore require more careful thought than area-weights. The second contribution of this paper is an exposition of the methods to deal with changing weights, developed in the discipline of economics. Specifically, I introduce the indices of \citet{laspeyres}, \citet{paasche} and \citet{fisher}.

I am not the first to depart from area-weighted trends in climate variables. For instance, \citet{Lobell616} and \citet{Auffhammer2013} compute cropping-area and growing-season trends in temperature and precipitation. However, they assume that cropping-area and growing season do not change. This implies that the estimated trend is representative only for part of the sample, and discrepancies grow with the distance to the base year. \citet{Dell2014} nonetheless recommend this method.

Studies of energy demand commonly use population-weighted temperatures, or rather population-weighted heating and cooling degree days. \citet{QuayleDiaz1980}, \citet{Taylor1981}, \citet{Guttman1983} and NOAA \footnote{See \href{http://www.cpc.ncep.noaa.gov/products/analysis_monitoring/cdus/degree_days/}{Degree Days Statistics}.} compute population-weighted degree-days, but with constant population-weights. These studies suffer the same problem as the Lobell paper cited above. On the other hand, \citet{Downton1988}, \citet{EIA2012} and \citet{Shi2016} compute population-weighted degree-days, with current population-weights. The resulting trend does not suffer from lack of representatives but\textemdash as made rigorous below\textemdash it does conflate the trend in temperature with the trend in the spatial distribution of the population. \citet{Michaels2014} fall into the same trap when computing the "experiential temperature" trend for the USA.

This paper presents a compromise between unrepresentative and conflated trends. The methods used are taken from price inflation. Price inflation measures the increase in average prices, say of consumer goods. This is a weighted average, as items more commonly bought should be more important. The common basket of goods changes over time and so the weights should be adjusted. Etienne Laspeyres and Hermann Paasche proposed methods to measure price inflation, later refined Irving Fisher. These methods have in common that the weights used ensure that the weighted average remains representative, and ensure that no spurious trend is introduced.

Obviously, the measurement of price inflation has moved on since the late 19\textsuperscript{th} century \citep{Diewert1998, Nordhaus1998, Hausman2003, Schultze2003, Broda2006, Handbury2015}. Most of the later refinements are in response to the peculiarities of consumer prices and expenditures, rather than about the general principles of trends in weighted averages. There are two issues particular to population-weighted temperature averages, viz. urbanization and migration. As a third contribution, I propose and apply methods to correct the temperature record for the urban heat island effect, and for international movement of people.

The paper proceeds as follows. Section \ref{sec:data} presents the data and their sources. Section \ref{sec:methods} discusses the methods for computing the population-weighted temperature trend in the presence of changes in population patterns (\ref{sec:popdens}), urbanization (\ref{sec:urban}), and international migration (\ref{sec:migrate}). Section \ref{sec:results} shows the results. Section \ref{sec:conclude} concludes.

\section{Data}
\label{sec:data}
Population data are taken from \citet{Goldewijk2010}. Data are population counts on a 0.5' $\times$ 0.5' grid. The Matlab code to convert the data to a 0.5\degree $\times$ 0.5\degree grid is given in the appendix. I use the data from 1900 to 2000, in ten year time steps. Unfortunately, the gridded population for 2010 has yet to be released. The data serve to illustrate the methods outlined below.

Temperature data are from CRU TS v3.24 \citep{Harris2014}. Data are monthly temperature average on a 0.5\degree $\times$ 0.5\degree grid. I use the data from 1900 to 2010, and compute the 21-year average centred on 1910, 1920, ..., 2000. The code is given in the appendix.

Data on urbanization are from the World Urbanization Prospects project of the UN Population Division.\footnote{see \href{http://esa.un.org/undp/wpup/CD-ROM}{Urban Agglomerations, file 12}} Specifically, the data are population counts, per pentade, from 1950 to 2015, for the 1692 urban agglomerations that had 300,000 inhabitants or more in 2014. \citet{Reba2016} has data that go back further in time, but coverage is not comprehensive.

The urban heat island effect is not consistently observed over space and time. Instead, I impute the warming due to urbanization by 
\begin{equation}
\label{eq:UHI}
U_{c,t} = \alpha P_{c,t}^{\beta}
\end{equation}
where
\begin{itemize}
\item $C_{c,t}$ is the warming due to the urban heat island effect in city $c$ at time $t$;
\item $P_{c,t}$ is the number of inhabitants of city $c$ at time $t$; and
\item $\alpha$ and $\beta$ are parameters.
\end{itemize}
Following \citet{Karl1988}, I set $\alpha=0.00174$ and $\beta=0.45$. These values are appropriate for annual temperatures for cities of over 100,000 people.

Data on international migration is taken from the UN Population Division,\footnote{see \href{http://www.un.org/en/development/desa/population/migration/data/estimates2/estimates15.shtml}{International migrant stock by destination and origin}.} which draws on \citet{Parsons2007}. This database has estimates of the total stock of migrants by origin and destination, per pentade, from 1990 to 2015. I am not aware of a global database that goes back further in time.

\section{Methods}
\label{sec:methods}
\subsection{Population density}
\label{sec:popdens}
As a first step, I compute the increase in the global annual mean surface air temperature weighted by the population per grid cell. Starting from the gridded temperature data, the global mean temperature is defined as
\begin{equation}
\label{eq:awt}
T_{t}^{A} := \frac{1}{\sum_{g=1}^G A_{g}} \sum_{g=1}^G T_{g,t} A_g
\end{equation}
where
\begin{itemize}
\item $T_{t}^{A}$ is the area-weighted global annual mean surface air temperature in year $t$;
\item $T_{g,t}$ is the annual mean surface air temperate in grid cell $g$ in year $t$; 
\item $A_g$ is the surface area of grid cell $g$; and
\item $G$ is the number of grid cells.
\end{itemize}
Because the grid cell area is constant over time, the change in the global mean temperature is defined as
\begin{equation}
\label{eq:dawt}
\Delta T_{t}^{A} := T_{t+1}^{A} - T_{t}^{A}
\end{equation}
The population-weighted temperature could be defined as in Equation (\ref{eq:awt}), with population-weights $P$ replacing area-weights $A$. However, as population is not constant over time and population growth not uniform over space, the equivalent of Equation (\ref{eq:dawt}) would conflate population and temperature change. Define
\begin{equation}
\label{eq:pop}
T_{t}^{P} := \frac{1}{\sum_{g=1}^G P_{g,t}} \sum_{g=1}^G T_{g,t} P_{g,t}
\end{equation}
where $P_{g,t}$ is the human population in grid cell $g$ in year $t$. Let $P_t$ be the total population at time $t$. Then
\begin{multline}
\label{eq:dpop}
\Delta T_{t}^{P} := T_{t+1}^{P} - T_{t}^{P} = \\
= \frac{1}{P_{t+1}} \sum_{g=1}^G T_{g,t+1} P_{g,t+1} -  \frac{1}{ P_{t}} \sum_{g=1}^G T_{g,t} P_{g,t} \approx \\
\approx   \sum_{g=1}^G \Delta T_{g,t} \frac{P_{g,t}}{P_{t}} + T_{g,t} \Delta \frac {P_{g,t}}{P_{t}}
\end{multline}
That is, Equation (\ref{eq:pop}) does not measure the temperature trend, but rather the temperature trend and the trend in the regional composition of the population.

The temperature change as experienced by the average person can be defined using the additive version of the chained Laspeyres Index \citep{laspeyres}
\begin{equation}
\label{eq:lpwt}
\Delta T_{t}^{P,L} :=\frac{1}{\sum_{g=1}^G P_{g,t}} \left ( \sum_{g=1}^G T_{g,t+1} P_{g,t} - \sum_{g=1}^G T_{g,t} P_{g,t} \right )
\end{equation}
where
\begin{itemize}
\item $T_{t}^{P,L}$ is the Laspeyres population-weighted global annual mean surface air temperature in year $t$; and
\item other variables are as above.
\end{itemize}

Equation (\ref{eq:lpwt}) uses population-weights for the start year. These weights are appropriate for that year, but less so for the end year.

Note that the Laspeyres Index in Equation (\ref{eq:lpwt}) defines the \textit{change} in the average temperature. The temperature level is found by starting from an arbitrary base (set at nought below) and adding subsequent changes.

Instead of the Laspeyres Index, we can use the chained Paasche Index \citep{paasche}
\begin{equation}
\label{eq:ppwt}
\Delta T_{t}^{P,P} :=\frac{1}{\sum_{g=1}^G P_{g,t+1}} \left ( \sum_{g=1}^G T_{g,t+1} P_{g,t+1} - \sum_{g=1}^G T_{g,t} P_{g,t+1} \right )
\end{equation}
or the Fisher Ideal Index \citep{fisher}
\begin{equation}
\label{eq:fpwt}
\Delta T_{t}^{P,F} :=0.5 \left (T_{t}^{P,L} + T_{t}^{P,P} \right )
\end{equation}
The Laspeyres index computes the temperature change as experienced by people in the base year $t$, the Paasche index as experienced by people in the new year $t+1$, and the Fisher ideal index takes the arithmetic average of the two. Laspeyres is representative for the past, Paasche for the present, and Fisher is a compromise between the two.

Equations (\ref{eq:lpwt}) and (\ref{eq:ppwt}) show the chained indices, in which the weights are updated every time step, but we can of course also use the population weights in the first (Laspeyres) and final period (Paasche).

The code to compute the various trends is given in the appendix.

\subsection{Urbanization}
\label{sec:urban}
Temperature records are homogenized, that is, temperature trends induced by changes in the equipment or the immediate environment of the weather station are removed\textemdash including the impact of urbanization. This makes perfect sense if the aim is to measure changes in the global environment, but not at all if the aim is to measure the warming experienced by humans. As a second step, therefore, I will reintroduce the urban heat island effect.

Let us redefine the population weighted global mean surface air temperature as
\begin{equation}
\label{eq:pwt}
T_{t}^{P^*} := \frac{1}{\sum_{g=1}^G P_{g,t}} \left [ \sum_{g=1}^G \left ( T_{g,t} +U_{g,t} \right ) P_{g,t} \nu_{g,t} + \sum_{g=1}^G T_{g,t} P_{g,t} \left ( 1 - \nu_{g,t} \right )\right ]
\end{equation}
where
\begin{itemize}
\item $U_{g,t}$ is the warming due to the urban heat island effect in grid cell $g$ at time $t$;
\item $\nu_{g,t}$ is the fraction of people living in urban centres in grid cell $g$ at time $t$; and
\item other variables are defined as above.
\end{itemize}
The Laspeyres index is then defined as
\begin{multline}
\label{eq:luwt}
\Delta T_{t}^{P^*,L} := \\
\frac{1}{\sum_{g=1}^G P_{g,t}} \left [ \sum_{g=1}^G \left ( T_{g,t+1} +U_{g,t+1} \right ) P_{g,t} \nu_{g,t} + \sum_{g=1}^G T_{g,t+1} P_{g,t} \left ( 1 - \nu_{g,t} \right ) \right. \\
\left. - \sum_{g=1}^G \left ( T_{g,t} +U_{g,t} \right ) P_{g,t} \nu_{g,t} - \sum_{g=1}^G T_{g,t} P_{g,t} \left ( 1 - \nu_{g,t} \right ) \right ] = \\
\Delta T_{t}^{P,L} + \frac{1}{\sum_{g=1}^G P_{g,t}} \left ( \sum_{g=1}^G U_{g,t+1} P_{g,t} \nu_{g,t} -  \sum_{g=1}^G U_{g,t} P_{g,t} \nu_{g,t} \right ) =: \\
\Delta T_{t}^{P,L} + \frac{\sum_{g=1}^G P_{g,t} \nu_{g,t}}{\sum_{g=1}^G P_{g,t}} \Delta U_{t}^{P,L}
\end{multline}

In other words, the Laspeyres index of global warming plus urban heat island is the Laspeyres index of global warming plus the Laspeyres index of urban heat island times the rate of urbanization. The same result carries over to the Paasche and Fisher indices.

\subsection{International Migration}
\label{sec:migrate}
People experience climate change not only in the place where they live, but also when they move. This may matter for population-weighted global warming because, although only relatively few people are affected, they experience relatively large climate change. This can be included as
\begin{equation}
\Delta T_{t}^{M} = \frac {1}{\sum_{o=1}^{O} \sum_{d=1}^{D} M_{o,d,t,t+1}}\sum_{o=1}^{O} \sum_{d=1}^{D} \left ( T_{d,t+1} - T_{o,t} \right ) M_{o,d,t,t+1}
\end{equation}
where $M_{o,d,t,t+1}$ denotes the number of people who migrated from $o$ to $d$ between time $t$ and $t+1$. This expression should be added to the population-weighted temperature after multiplication by the share of people migrating.

\section{Results}
\label{sec:results}
\subsection{Population density}
\begin{figure}[h]
\includegraphics[width=\textwidth]{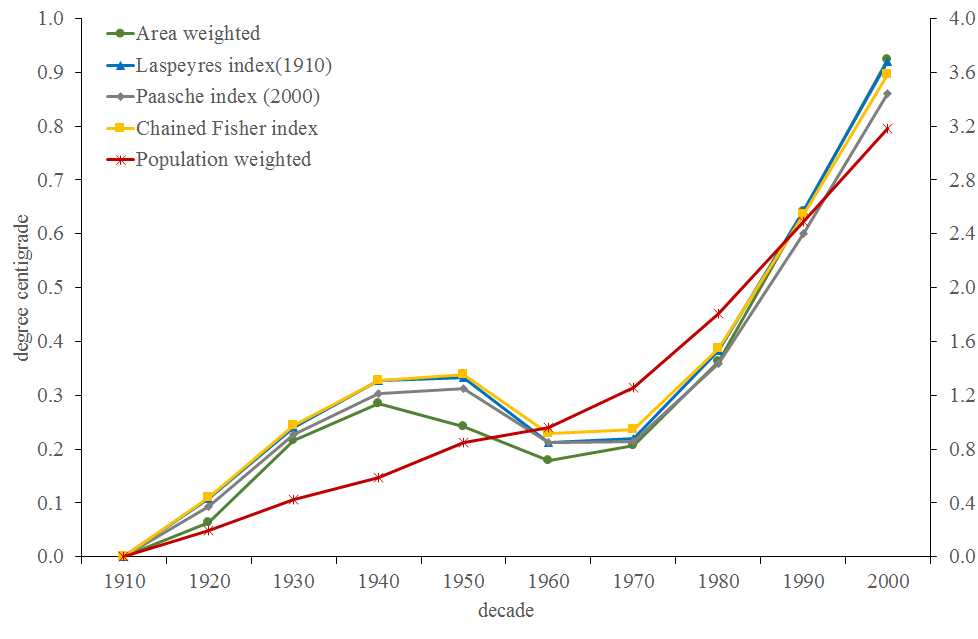}
\caption{The global mean surface air temperature over land in degrees Celsius, weighted by area and weighted by population size, using the Laspeyres Index, the Paasche Index, and the chained Fisher Index (left axis) and weighted by current population (right axis).}
\label{fig:indextemp}
\end{figure}

Figure \ref{fig:indextemp} shows the global mean surface air temperature over land for four alternative methods. All temperature records are shown as anomalies from 1910. The four records show the same overall trend\textemdash a 0.9\celsius\: warming over the century\textemdash but the details are different: The population-weighted records warm faster than the area-weighted record until 1950, and slower afterwards. The difference between the area-weighted temperature and the chained Fisher index is 0.1\celsius\: in 1950; and this gap closes again by 2000. These differences matter, say, if you would regress economic growth on climate change, because the slope of the Fisher index is 40\% steeper in the first half of the century, and 18\% shallower in the second half.

The differences between the population-weighted temperature records are less pronounced. If we follow Laspeyres and use the 1910 population as weights, overall warming is 0.06\celsius\: higher than if we follow Paasche and use the 2000 population. The Fisher index lies in between. The pattern of warming until 1940, cooling until 1970, and warming after 1970 is the same in all three records, but the slopes differ. For instance, mid-century cooling and late-century warming are most pronounced in the Laspeyres index.

For comparison, Figure \ref{fig:indextemp} also shows the population-weighted temperature trend if current population weights are used, as in Equations (\ref{eq:pop}) and (\ref{eq:dpop}). The difference is large. Warming is now 3.2\celsius\: over the century. As argued above, this approach conflates the trend in the temperature with the trend in the distribution of the population, and the latter clearly dominates.

\subsection{Urbanization}
\begin{figure}[h]
\includegraphics[width=\textwidth]{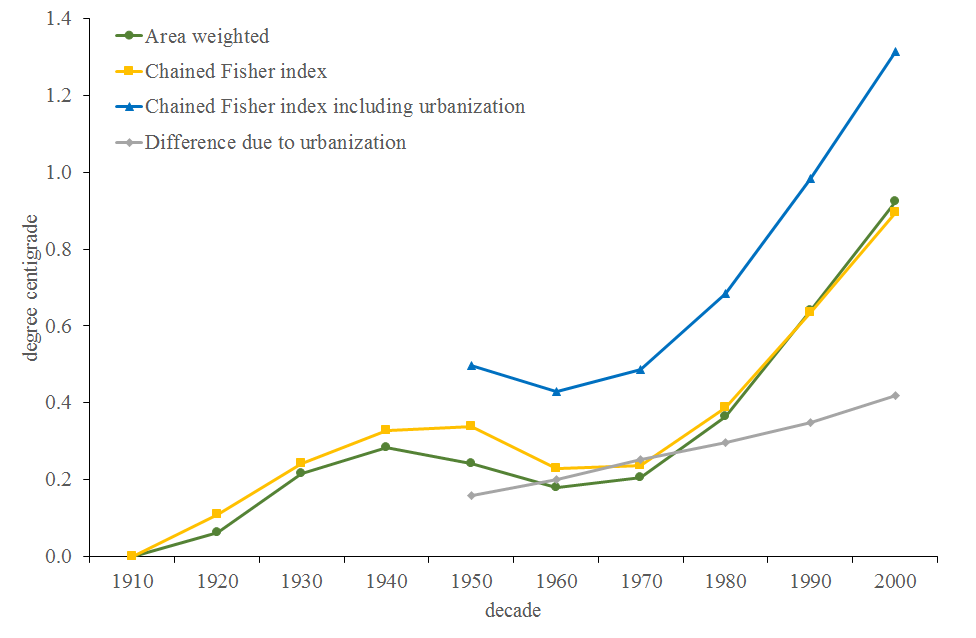}
\caption{The global mean surface air temperature over land in degrees Celsius, weighted by area and weighted by population size, using the chained Fisher Index, with and without urbanization. The difference due to urbanization is also shown.}
\label{fig:urban}
\end{figure}

Figure \ref{fig:urban} shows the impact of urbanization and the urban heat island effect on the population-weighted average temperature. In 1950, the average urban heat island effect is 0.98\celsius. This rises to 1.64\celsius\: in the year 2000. In 1950, 16\% of the world population lived in the cities for which we have data, rising to 26\% in 2000. Urbanization thus adds 0.16\celsius\: to the temperature in 1950 and 0.42\celsius in 2000. As the Fisher index rose by 0.54\celsius\: between 1950 and 2000, the extra warming of 0.26\celsius\: due to urbanization is relatively large.

\subsection{International migration}
The first estimate of the stock of international migrants by country of origin and destination is for the year 1990. In that year, 154 million migrants lived in countries 1.60\celsius\: colder, on average, than their county of origin. By the year 2000, there were 175 million migrants living in countries on average 2.85\celsius\: colder than where they came from. That is, in the last decade of the 20\textsuperscript{th} century, 20 million migrants experienced a cooling of 12.3\celsius. This is large relative to the observed warming, but additional migrants are only some 0.4\% of the world population, so the overall experience of warming does not change much: We need to substract 0.04\celsius\: from the temperatures shown in Figure \ref{fig:urban}.

\begin{figure}[h]
\includegraphics[width=\textwidth]{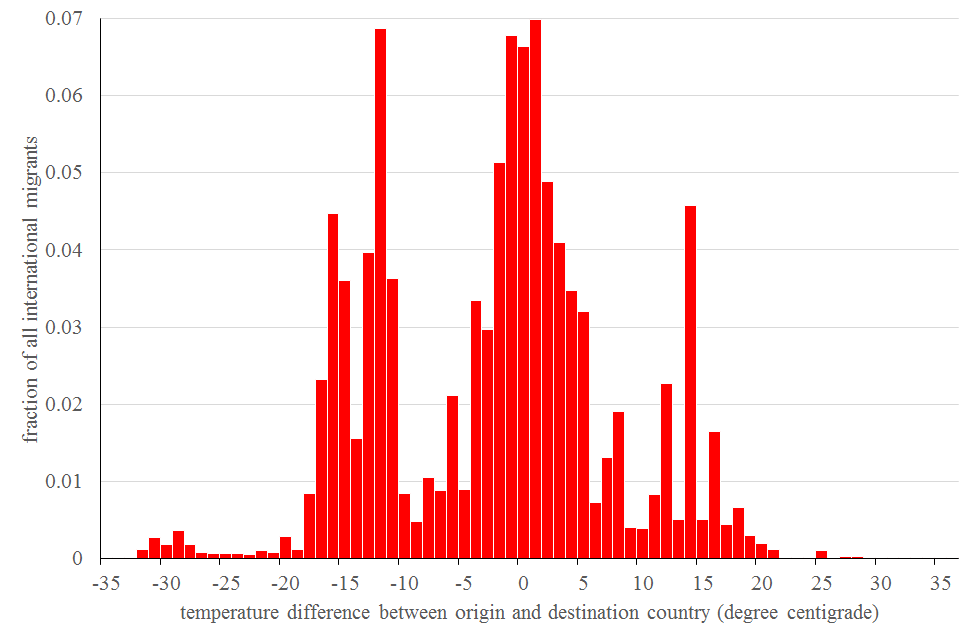}
\caption{Histogram of the warming or cooling experienced by the migrant population in the year 2000.}
\label{fig:migration}
\end{figure}

Figure \ref{fig:migration} shows the range of temperature change as experienced by those who moved country in 2000 or before. These changes are large. A quarter of migrants experienced warming or cooling of 2\celsius\: or less, but 42\% experienced an absolute temperature change of 10\celsius\: or more. 60\% of migrants moved from warmer to colder countries. The 40\% who experienced warming, and particularly the 28\% who experienced warming of more than 2\celsius\: could be used as case studies of the impacts of future climate change.

\section{Discussion and conclusion}
\label{sec:conclude}
Previous studies applying population-weighted average temperatures used either constant weights, so that the average is not representative for part of the sample, or current weights, so that the trend conflates trends in the temperature with trends in the distribution of the population. I propose to use the methods used to estimate price inflation, particularly the Fisher Ideal Index, to derive population-weighted trends in temperature as this maintains representativeness without conflation. I extend these methods to include urbanization and the urban heat island effect, and to reflect international migration. An application to the temperature trends in the 20\textsuperscript{th} century shows (1) that the population-weighted temperature trend differs in key details from the area-weighted temperature trend; (2) that conflation is a bigger problem than non-representativeness; (3) that the urban heat island effect substantially adds to warming; and (4) that international migration lead to a minor drop in the temperature experienced by the average person.

The methods proposed here can be used for any climate variable and for any weight. Temperature and population are just one of many possible applications. Rainfall and cropping area would be an obvious next application.

Empirical papers that relate changes in climate to changes in economic activity \citep{Dell2012, Burke2015, Lemoine2015}, crop yields \citep{Lobell2003, Chen2004, Lin2012}, species abundance \citep{Root2003, Parmesan2003, Atkins2010, Eglington2012} or some other variable of interest should take care in constructing the climate variable of interest. As shown above, the slope differs between area- and population-weighted temperature trends. This could well lead to spurious inference. Recall that the area- and population-weighted temperatures diverge in the first half of the 20\textsuperscript{th} century and converge in the second. An impact indicator that co-trends with either can therefore not co-trend with the other.

There is thus a rich vein of future research following on from the methods proposed here. The paper also identifies the experience of migration as a laboratory for the impact of climate change.

\bibliography{tol}

\section{Appendix: Matlab codes}

\ttfamily

\subsection{processarea.m}

load gridarea.txt \\
\\
\%area for 5'x5' grid\\
gridarea(gridarea==-9999)=0;\\
\\
area = zeros(NLat, NLong);\\
\\
\%aggregate 5'x5' grid to 30'x30' grid\\
for i = 1:NLat\\
\hspace*{0.5cm} for j = 1:NLong\\
\hspace*{1.0cm} for k=1:6\\
\hspace*{1.5cm} for l=1:6\\
\hspace*{2.0cm} area(i,j) = area(i,j) + gridarea(k+(i-1)*6,l+(j-1)*6);\\
\hspace*{1.5cm} end\\
\hspace*{1.0cm} end\\
\hspace*{0.5cm} end\\
end\\

\subsection{processHYDE.m}

clear all\\
\\
processarea\\
\\
\%load HYDE data\\
load popc1900AD.txt\\
load popc1910AD.txt\\
load popc1920AD.txt\\
load popc1930AD.txt\\
load popc1940AD.txt\\
load popc1950AD.txt\\
load popc1960AD.txt\\
load popc1970AD.txt\\
load popc1980AD.txt\\
load popc1990AD.txt\\
load popc2000AD.txt\\
\\
popc1900AD(popc1900AD==-9999)=0;\\
popc1910AD(popc1910AD==-9999)=0;\\
popc1920AD(popc1920AD==-9999)=0;\\
popc1930AD(popc1930AD==-9999)=0;\\
popc1940AD(popc1940AD==-9999)=0;\\
popc1950AD(popc1950AD==-9999)=0;\\
popc1960AD(popc1960AD==-9999)=0;\\
popc1970AD(popc1970AD==-9999)=0;\\
popc1980AD(popc1980AD==-9999)=0;\\
popc1990AD(popc1990AD==-9999)=0;\\
popc2000AD(popc2000AD==-9999)=0;\\
\\
NLong = 720;\\
NLat = 360;\\
\\
P1900 = zeros(NLat, NLong);\\
P1910 = zeros(NLat, NLong);\\
P1920 = zeros(NLat, NLong);\\
P1930 = zeros(NLat, NLong);\\
P1940 = zeros(NLat, NLong);\\
P1950 = zeros(NLat, NLong);\\
P1960 = zeros(NLat, NLong);\\
P1970 = zeros(NLat, NLong);\\
P1980 = zeros(NLat, NLong);\\
P1990 = zeros(NLat, NLong);\\
P2000 = zeros(NLat, NLong);\\
\\
\%aggregate 5'x5' grid to 30'x30' grid\\
for i = 1:NLat\\
\hspace*{0.5cm} for j = 1:NLong\\
\hspace*{1.0cm} for k=1:6\\
\hspace*{1.5cm} for l=1:6\\
\hspace*{2.0cm} P1900(i,j) = P1900(i,j) + popc1900AD(k+(i-1)*6,l+(j-1)*6);\\
\hspace*{2.0cm} P1910(i,j) = P1910(i,j) + popc1910AD(k+(i-1)*6,l+(j-1)*6);\\
\hspace*{2.0cm} P1920(i,j) = P1920(i,j) + popc1920AD(k+(i-1)*6,l+(j-1)*6);\\
\hspace*{2.0cm} P1930(i,j) = P1930(i,j) + popc1930AD(k+(i-1)*6,l+(j-1)*6);\\
\hspace*{2.0cm} P1940(i,j) = P1940(i,j) + popc1940AD(k+(i-1)*6,l+(j-1)*6);\\
\hspace*{2.0cm} P1950(i,j) = P1950(i,j) + popc1950AD(k+(i-1)*6,l+(j-1)*6);\\
\hspace*{2.0cm} P1960(i,j) = P1960(i,j) + popc1960AD(k+(i-1)*6,l+(j-1)*6);\\
\hspace*{2.0cm} P1970(i,j) = P1970(i,j) + popc1970AD(k+(i-1)*6,l+(j-1)*6);\\
\hspace*{2.0cm} P1980(i,j) = P1980(i,j) + popc1980AD(k+(i-1)*6,l+(j-1)*6);\\
\hspace*{2.0cm} P1990(i,j) = P1990(i,j) + popc1990AD(k+(i-1)*6,l+(j-1)*6);\\
\hspace*{2.0cm} P2000(i,j) = P2000(i,j) + popc2000AD(k+(i-1)*6,l+(j-1)*6);\\
\hspace*{1.5cm} end\\
\hspace*{1.0cm} end\\
\hspace*{0.5cm} end\\
end\\
\\
save population\\

\subsection{processCRU.m}
clear all\\
\\
NLong = 720;\\
NLat = 360;\\
NMonth = 12;\\
NYear = 113;\\
NLines = NLat*NMonth*NYear;\\
\\
load cru\_ ts3.22.1901.2013.tmp.dat\\
A=cru\_ ts3\_ 22\_ 1901\_ 2013\_ tmp;\\
clear cru*\\
A(A==-999)=NaN;\\
\\
B = zeros(NLat*NYear, NLong);\\
\\
\%compute annual average temperature\\
for y = 1:NYear\\
\hspace*{0.5cm} for m = 1:NMonth\\
\hspace*{1.0cm} for l = 1:NLat\\
\hspace*{1.5cm} Aind = l+(m-1)*NLat+(y-1)*NLat*NMonth;\\
\hspace*{1.5cm} Bind = l+(y-1)*NLat;\\
\hspace*{1.5cm} B(Bind,:) = B(Bind,:) + A(Aind,:)/12;\\
\hspace*{1.0cm} end\\
\hspace*{0.5cm} end\\
end\\
\\
T1910 = zeros(NLat,NLong);\\
T1920 = zeros(NLat,NLong);\\
T1930 = zeros(NLat,NLong);\\
T1940 = zeros(NLat,NLong);\\
T1950 = zeros(NLat,NLong);\\
T1960 = zeros(NLat,NLong);\\
T1970 = zeros(NLat,NLong);\\
T1980 = zeros(NLat,NLong);\\
T1990 = zeros(NLat,NLong);\\
T2000 = zeros(NLat,NLong);\\
\\
NAve = 21;\\
\\
\%compute 21-year average temperature and store it in a grid\\
for y = 1:NAve\\
\hspace*{0.5cm} for l = 1:NLat\\
\hspace*{1.0cm} T1910(l,:) = T1910(l,:) + B(l + (y-1)*NLat,:)/NAve;\\
\hspace*{1.0cm} T1920(l,:) = T1920(l,:) + B(l + (y+10-1)*NLat,:)/NAve;\\
\hspace*{1.0cm} T1930(l,:) = T1930(l,:) + B(l + (y+20-1)*NLat,:)/NAve;\\
\hspace*{1.0cm} T1940(l,:) = T1940(l,:) + B(l + (y+30-1)*NLat,:)/NAve;\\
\hspace*{1.0cm} T1950(l,:) = T1950(l,:) + B(l + (y+40-1)*NLat,:)/NAve;\\
\hspace*{1.0cm} T1960(l,:) = T1960(l,:) + B(l + (y+50-1)*NLat,:)/NAve;\\
\hspace*{1.0cm} T1970(l,:) = T1970(l,:) + B(l + (y+60-1)*NLat,:)/NAve;\\
\hspace*{1.0cm} T1980(l,:) = T1980(l,:) + B(l + (y+70-1)*NLat,:)/NAve;\\
\hspace*{1.0cm} T1990(l,:) = T1990(l,:) + B(l + (y+80-1)*NLat,:)/NAve;\\
\hspace*{1.0cm} T2000(l,:) = T2000(l,:) + B(l + (y+90-1)*NLat,:)/NAve;\\
\hspace*{0.5cm} end\\
end\\
\\
save temperature\\

\subsection{processall.m}
clear all\\
\\
load temperature\\
load population\\
\\
totarea = sum(sum(area));\\
\\
\%divide by ten to convert from milligrade to centigrade, add 273 to convert from Celsius to Kelvin 
T1910 = T1910/10 + 273.15;\\
T1920 = T1920/10 + 273.15;\\
T1930 = T1930/10 + 273.15;\\
T1940 = T1940/10 + 273.15;\\
T1950 = T1950/10 + 273.15;\\
T1960 = T1960/10 + 273.15;\\
T1970 = T1970/10 + 273.15;\\
T1980 = T1980/10 + 273.15;\\
T1990 = T1990/10 + 273.15;\\
T2000 = T2000/10 + 273.15;\\
\\
gTemp = (T2000-T1910)./T1910;\\
gTemp(isnan(gTemp)) = 0;\\
\\
T1910(isnan(T1910)) = 0;\\
T1920(isnan(T1920)) = 0;\\
T1930(isnan(T1930)) = 0;\\
T1940(isnan(T1940)) = 0;\\
T1950(isnan(T1950)) = 0;\\
T1960(isnan(T1960)) = 0;\\
T1970(isnan(T1970)) = 0;\\
T1980(isnan(T1980)) = 0;\\
T1990(isnan(T1990)) = 0;\\
T2000(isnan(T2000)) = 0;\\
\\
area = flipud(area);\\
\\
\%area-weighted temperature
GMT(1) = sum(sum(T1910.*area))/totarea;\\
GMT(2) = sum(sum(T1920.*area))/totarea;\\
GMT(3) = sum(sum(T1930.*area))/totarea;\\
GMT(4) = sum(sum(T1940.*area))/totarea;\\
GMT(5) = sum(sum(T1950.*area))/totarea;\\
GMT(6) = sum(sum(T1960.*area))/totarea;\\
GMT(7) = sum(sum(T1970.*area))/totarea;\\
GMT(8) = sum(sum(T1980.*area))/totarea;\\
GMT(9) = sum(sum(T1990.*area))/totarea;\\
GMT(10) = sum(sum(T2000.*area))/totarea;\\
\\
P1910 = flipud(P1910);\\
P1920 = flipud(P1920);\\
P1930 = flipud(P1930);\\
P1940 = flipud(P1940);\\
P1950 = flipud(P1950);\\
P1960 = flipud(P1960);\\
P1970 = flipud(P1970);\\
P1980 = flipud(P1980);\\
P1990 = flipud(P1990);\\
P2000 = flipud(P2000);\\
\\
gPop = (P2000-P1910)./P1910;\\
gPop(isnan(gPop)) = 0;\\
\\
\%Laspeyres\\
GMTL(1) = sum(sum(T1910.*P1910))/sum(sum(P1910));\\
GMTL(2) = sum(sum(T1920.*P1910))/sum(sum(P1910));\\
GMTL(3) = sum(sum(T1930.*P1910))/sum(sum(P1910));\\
GMTL(4) = sum(sum(T1940.*P1910))/sum(sum(P1910));\\
GMTL(5) = sum(sum(T1950.*P1910))/sum(sum(P1910));\\
GMTL(6) = sum(sum(T1960.*P1910))/sum(sum(P1910));\\
GMTL(7) = sum(sum(T1970.*P1910))/sum(sum(P1910));\\
GMTL(8) = sum(sum(T1980.*P1910))/sum(sum(P1910));\\
GMTL(9) = sum(sum(T1990.*P1910))/sum(sum(P1910));\\
GMTL(10) = sum(sum(T2000.*P1910))/sum(sum(P1910));\\
\\
\%Paasche\\
GMTP(1) = sum(sum(T1910.*P2000))/sum(sum(P2000));\\
GMTP(2) = sum(sum(T1920.*P2000))/sum(sum(P2000));\\
GMTP(3) = sum(sum(T1930.*P2000))/sum(sum(P2000));\\
GMTP(4) = sum(sum(T1940.*P2000))/sum(sum(P2000));\\
GMTP(5) = sum(sum(T1950.*P2000))/sum(sum(P2000));\\
GMTP(6) = sum(sum(T1960.*P2000))/sum(sum(P2000));\\
GMTP(7) = sum(sum(T1970.*P2000))/sum(sum(P2000));\\
GMTP(8) = sum(sum(T1980.*P2000))/sum(sum(P2000));\\
GMTP(9) = sum(sum(T1990.*P2000))/sum(sum(P2000));\\
GMTP(10) = sum(sum(T2000.*P2000))/sum(sum(P2000));\\
\\
\%population\\
GMTp(1) = sum(sum(T1910.*P1910))/sum(sum(P1910));\\
GMTp(2) = sum(sum(T1920.*P1920))/sum(sum(P1920));\\
GMTp(3) = sum(sum(T1930.*P1930))/sum(sum(P1930));\\
GMTp(4) = sum(sum(T1940.*P1940))/sum(sum(P1940));\\
GMTp(5) = sum(sum(T1950.*P1950))/sum(sum(P1950));\\
GMTp(6) = sum(sum(T1960.*P1960))/sum(sum(P1960));\\
GMTp(7) = sum(sum(T1970.*P1970))/sum(sum(P1970));\\
GMTp(8) = sum(sum(T1980.*P1980))/sum(sum(P1980));\\
GMTp(9) = sum(sum(T1990.*P1990))/sum(sum(P1990));\\
GMTp(10) = sum(sum(T2000.*P2000))/sum(sum(P2000));\\
\\
\%chained Laspeyres\\
CL(1) = 0;\\
CL(2) = sum(sum(T1920.*P1910))/sum(sum(P1910)) - sum(sum(T1910.*P1910))/sum(sum(P1910));\\
CL(3) = sum(sum(T1930.*P1920))/sum(sum(P1920)) - sum(sum(T1920.*P1920))/sum(sum(P1920));\\
CL(4) = sum(sum(T1940.*P1930))/sum(sum(P1930)) - sum(sum(T1930.*P1930))/sum(sum(P1930));\\
CL(5) = sum(sum(T1950.*P1940))/sum(sum(P1940)) - sum(sum(T1940.*P1940))/sum(sum(P1940));\\
CL(6) = sum(sum(T1960.*P1950))/sum(sum(P1950)) - sum(sum(T1950.*P1950))/sum(sum(P1950));\\
CL(7) = sum(sum(T1970.*P1960))/sum(sum(P1960)) - sum(sum(T1960.*P1960))/sum(sum(P1960));\\
CL(8) = sum(sum(T1980.*P1970))/sum(sum(P1970)) - sum(sum(T1970.*P1970))/sum(sum(P1970));\\
CL(9) = sum(sum(T1990.*P1980))/sum(sum(P1980)) - sum(sum(T1980.*P1980))/sum(sum(P1980));\\
CL(10) = sum(sum(T2000.*P1990))/sum(sum(P1990)) - sum(sum(T1990.*P1990))/sum(sum(P1990));\\
\\
\%chained Paasche\\
CP(1) = 0;\\
CP(2) = sum(sum(T1920.*P1920))/sum(sum(P1920)) - sum(sum(T1910.*P1920))/sum(sum(P1920));\\
CP(3) = sum(sum(T1930.*P1930))/sum(sum(P1930)) - sum(sum(T1920.*P1930))/sum(sum(P1930));\\
CP(4) = sum(sum(T1940.*P1940))/sum(sum(P1940)) - sum(sum(T1930.*P1940))/sum(sum(P1940));\\
CP(5) = sum(sum(T1950.*P1950))/sum(sum(P1950)) - sum(sum(T1940.*P1950))/sum(sum(P1950));\\
CP(6) = sum(sum(T1960.*P1960))/sum(sum(P1960)) - sum(sum(T1950.*P1960))/sum(sum(P1960));\\
CP(7) = sum(sum(T1970.*P1970))/sum(sum(P1970)) - sum(sum(T1960.*P1970))/sum(sum(P1970));\\
CP(8) = sum(sum(T1980.*P1980))/sum(sum(P1980)) - sum(sum(T1970.*P1980))/sum(sum(P1980));\\
CP(9) = sum(sum(T1990.*P1990))/sum(sum(P1990)) - sum(sum(T1980.*P1990))/sum(sum(P1990));\\
CP(10) = sum(sum(T2000.*P2000))/sum(sum(P2000)) - sum(sum(T1990.*P2000))/sum(sum(P2000));\\
\\
\%chained Fisher\\
CF = 0.5*(CL + CP);\\
\\
GMTCL(1) = GMTL(1);\\
GMTCP(1) = GMTL(1);\\
GMTCF(1) = GMTL(1);\\
\\
for i=2:10\\
\hspace*{0.5cm} GMTCL(i) = GMTCL(i-1) + CL(i);\\
\hspace*{0.5cm} GMTCP(i) = GMTCP(i-1) + CP(i);\\
\hspace*{0.5cm} GMTCF(i) = GMTCF(i-1) + CF(i);\\
end\\
\\
GMT   = GMT - GMT(1);\\
GMTL = GMTL - GMTL(1);\\
GMTP = GMTP - GMTP(1);\\
GMTp  = GMTp - GMTp(1);\\
GMTCL = GMTCL - GMTCL(1);\\
GMTCP = GMTCP - GMTCP(1);\\
GMTCF = GMTCF - GMTCF(1);\\
\\
dec = [1910 1920 1930 1940 1950 1960 1970 1980 1990 2000];\\
\\
plot(dec,GMT,dec,GMTL,dec,GMTP,dec,GMTCF),legend('area','pop1910','pop2000','Chained Fisher')\\

\end{document}